\documentstyle[11pt,newpasp,epsf]{article}

\def\etal{\it et~al.\rm\ }
\def\eg{e.g.,\ }

\begin{document}

\title{Non-equilibrium Kinematics in Merging Galaxies}
\author{J.C. Mihos}
\affil{Dept. of Astronomy, Case Western Reserve University, Cleveland, OH}

\begin{abstract}
Measurements of the kinematics of merging galaxies are often used to derive
dynamical masses, study evolution onto the fundamental plane, or probe
relaxation processes. These measurements are often compromised to some degree
by strong non-equilibrium motions in the merging galaxies. This talk focuses
on the evolution of the kinematics of merging galaxies, and highlights some
pitfalls which occur when studying non-equilibrium systems.

\end{abstract}


\section{Evolution of Velocity Moments in Merging Galaxies}

The global kinematics of merging galaxies are often used to infer dynamical
masses, or study evolution of merger remnants onto the fundamental plane
(\eg Lake \& Dressler 1986; Shier \etal 1994; James \etal 1999). In systems 
well out of equilibrium, these measurements 
may not yield true estimates of the velocity dispersion of the system.
For example, in a merger where the nuclei have not yet coalesced, much of the 
kinetic energy of the the system may be in bulk motion of the nuclei, rather
than in pure random stellar motions. Such conditions could in principle lead
to systematic errors in dynamical masses or fundamental plane properties. 
Equally important is the timescale over which any merger-induced kinematic 
irregularities are mixed away through violent relaxation or mixing.

To examine the evolution of the kinematic moments of a galaxy merger, Figure
1 shows the projected velocity moments in an N-body model of an equal mass 
galaxy merger. The data is constructed to simulate observations with modest
spatial resolution of $\sim$ 1 kpc.
The low order moments of the velocity distribution very quickly evolve to
their final value -- violent relaxation in the inner regions is 
extremely efficient. Even during the final coalescence phase, the velocity
dispersion of the merger is essentially unchanging, except for extreme
situations where the remnant is viewed almost exactly along the orbital
plane. This analysis suggests that studies which place mergers on the 
fundamental plane are not excessively compromised by possible kinematic
evolution of the remnants; instead, luminosity evolution should dominate
any changes in the properties of the remnant. 

\begin{figure}
\plotfiddle{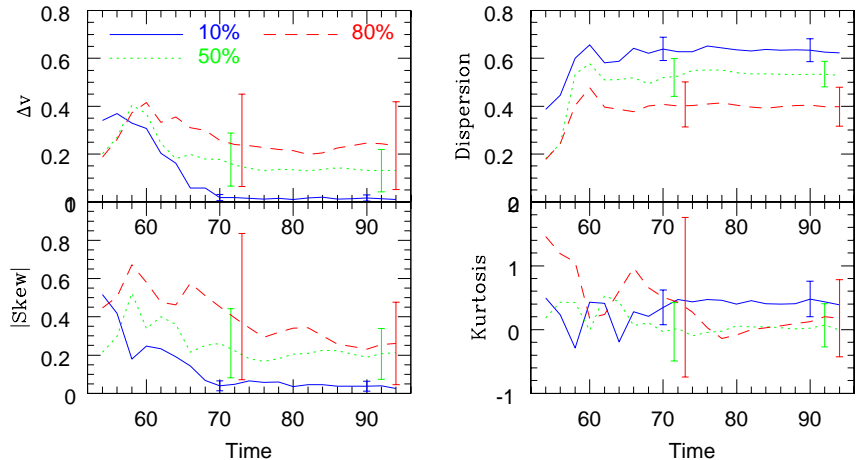}{2.0truein}{0}{60}{60}{-180}{-250}
\caption{Evolution of the projected velocity moments during an equal mass
merger. The three curves show local measurements at radii containing 10\%,
50\%, and 80\% of the total stellar mass of the system; ``error bars'' show
the variance of the measured moments due to viewing geometry. $\Delta v$
is the velocity difference between the opposite sides of the galaxies.  The
nuclei are separated by $\sim$ 5 kpc at T=60, and coalesce at T=70. To scale to
Milky Way-type progenitors, unit time is 13 Myr and unit velocity is 250
km/s.}
\end{figure}

At larger radius, the merger 
remnant possesses a significant rotational component, as transfer of
orbital angular momentum has spun up the remnant (\eg Hernquist 1992).
The higher order velocity moments (skew and kurtosis) continue to evolve for
several dynamical times, particularly in the outer portions of the remnant
where the mixing timescale is long. These higher order moments also vary
significantly with viewing angle, reflecting the fact that the merger
kinematics maintain a ``memory'' of the initial orbital angular momentum.
As high angular momentum material streams back into the remnant from the 
tidal debris, incomplete mixing results in extremely non-gaussian line
profiles. 

\section{Local Stellar Kinematics and Ghost Masses}

On smaller scales, however, measurements of local velocity dispersion can
give erroneous results if the system has not yet relaxed. Figure 2 shows
the merger model ``observed'' at higher spatial resolution at a time
when the nuclei are still separated by a few kpc. Looking along the orbital
plane, the nuclei still possess a significant amount of bulk motion.
Measured on small scales, this bulk motion shows as a gradient in the
projected radial velocity across the two nuclei. Perhaps more interesting
is the rise in projected velocity dispersion between the nuclei, where
the the velocity profile shows a single broad line  with dispersion
$\sim$ 30\% higher than in the nuclei themselves.  A similar rise is
seen between the nuclei of NGC 6240 (Tezca \etal 1999 in prep, referenced 
in Tacconi \etal 1999), where a central gas concentration exists.
The simulations here indicate that such features can arise in double nucleus
systems even when no central mass exists, and suggest that dynamical masses
inferred this way can be significantly overestimated.

In this case, the full analysis of the line profiles results in a better 
understanding of the dynamical conditions. The gradient across the nuclei
again is an indicator of large bulk motions, and the shape of the line
profile is rather flat-topped (negative kurtosis), exactly what is expected
from the incomplete blending of two separate line profiles. Here, of course,
the increase in velocity dispersion is due simply to the projected overlap
of the nuclei, but the complete line profile is needed to unravel the
complex dynamics. 

\section{Ionized Gas Kinematics and Starburst Winds}

Finally, while gas kinematics are perhaps the easiest to measure, 
they give the most ambiguous
measurement of the gravitational kinematics of a merging system. Aside from
the problems of the evolving gravitational kinematics and
line-of-sight projection effects, gas kinematics are also subject to 
influences such as shocks, radial inflow, and starburst winds. All of these
conspire to make a very confusing kinematic dataset.

\begin{figure}
\plotone{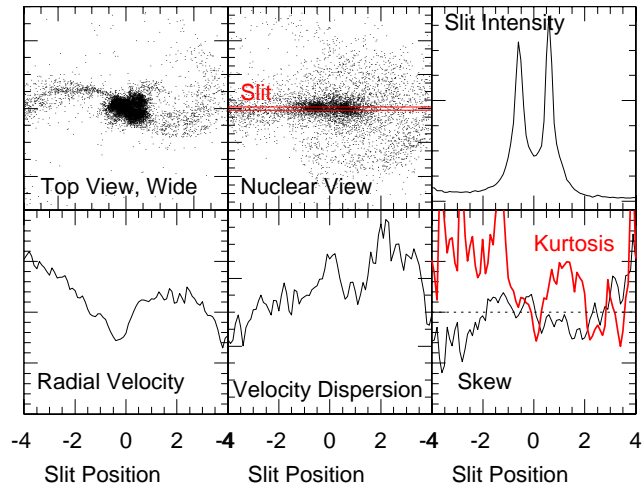}
\caption{Simulated longslit velocity moments for an unrelaxed merger; the
nuclei are in the final stages of coalescence. Unit length is $\sim$ 3.5 kpc.}
\end{figure}

A case in point is the ultraluminous infrared galaxy NGC 6240. This 
starburst system has a double nucleus separated by $\sim$ 1.5\arcsec\ and
is clearly a late stage merger. Based on H$\alpha$ velocity mapping of 
this system, Bland-Hawthorn \etal (1991) proposed that a $10^{12} M_{\sun}$
black hole exists well outside the nucleus, at a projected distance of
6 kpc. The major piece of evidence supporting this claim was a sharp
gradient in the ionized gas kinematics, suggestive of a rapidly rotating
disk.

To study this object in more detail, we (van der Marel \etal in prep)
have initiated a program using HST to obtain imaging and longslit 
spectroscopic data for the inner regions of NGC 6240. Figure 3 shows an F814W
image of the center of NGC 6240, along with a narrow band image centered on
H$\alpha$+[NII] (taken using the F673N filter, which for NGC 6240 fortuitously
sits on redshifted H$\alpha$).  The narrow band image shows a clear starburst 
wind morphology in the ionized gas.

\begin{figure}
\plotone{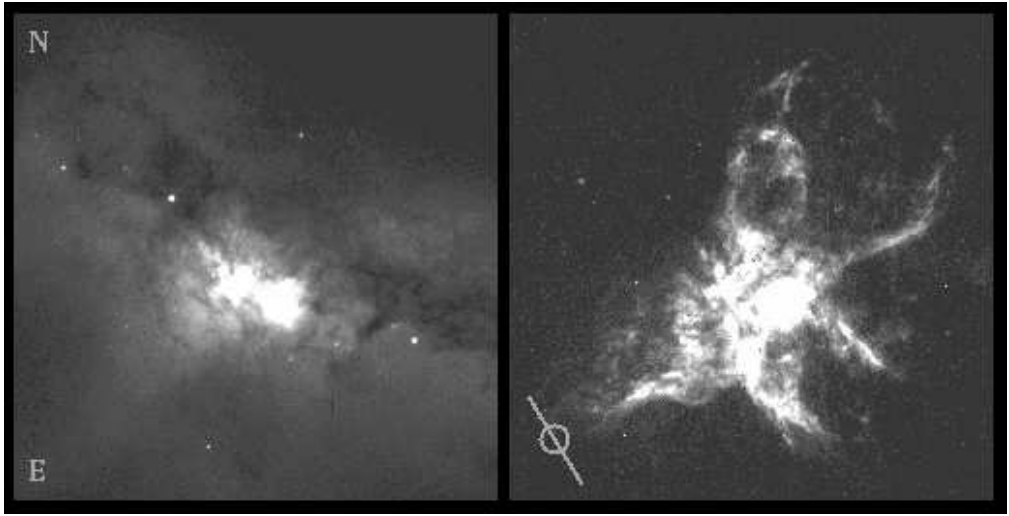}
\caption{Left: F814W image of the center of NGC 6240. Right: Narrow band
H$\alpha$ image. The position of the reported $10^{12} M_{\sun}$ black
hole (Bland-Hawthorn \etal 1991) is shown, along with the position angle 
of the purported disk surrounding it.}
\end{figure}

Overplotted on Figure 3b is the position of the putative black hole, along
with the position angle of the observed velocity gradient. Interestingly,
the position lies directly along an ionized filament from the starburst
wind, with the kinematic gradient directed orthogonal to the filament's
direction. While our narrow-band data do not go deep enough to study the
detailed distribution of ionized gas immediately surrounding the proposed
black hole, the image certainly suggests that the observed kinematics
may be strongly influenced by the starburst wind, indicating that the
black hole may not be real. The strong gradient that was attributed to a
black hole may instead be due to kinematic gradients in the starburst wind, 
or even simple geometry of the wind filament projecting on top 
of background system emission.  We have follow-up STIS spectroscopy
planned to further study the complex kinematics in this intriguing system.

\acknowledgments

This work was sponsored in part by the San Diego Supercomputing Center,
the NSF, and STScI. I thank Rebecca Stanek and Sean Maxwell for help with
data analysis.


\begin{references}
\reference Bland-Hawthorn, J., Wilson, A.S., \& Tully, R.B. 1991, \apj, 371, L19.
\reference Hernquist, L. 1992, \apj, 400, 460
\reference James, P., \etal, astro-ph/9906276
\reference Lake, G., \& Dressler, A. 1986, \apj, 310, 605
\reference Shier, L.M., Rieke, M.J., \& Rieke, G.H. 1994, \apj, 433, L9
\reference Tacconi, L.J., \etal, astro-ph/9905031
\reference 
\end{references}
\end{document}